\documentclass[twocolumn,showpacs,preprintnumbers,amsmath,amssymb,superscriptaddress]{revtex4}

\usepackage{graphicx}
\usepackage{dcolumn}
\usepackage{bm}

\begin{document}

\preprint{APS/123-QED}

\title{Limits on a muon flux from neutralino annihilations in the Sun\\with the IceCube 22-string detector}

\affiliation{III Physikalisches Institut, RWTH Aachen University, D-52056 Aachen, Germany}
\affiliation{Dept.~of Physics and Astronomy, University of Alabama, Tuscaloosa, AL 35487, USA}
\affiliation{Dept.~of Physics and Astronomy, University of Alaska Anchorage, 3211 Providence Dr., Anchorage, AK 99508, USA}
\affiliation{CTSPS, Clark-Atlanta University, Atlanta, GA 30314, USA}
\affiliation{School of Physics and Center for Relativistic Astrophysics, Georgia Institute of Technology, Atlanta, GA 30332. USA}
\affiliation{Dept.~of Physics, Southern University, Baton Rouge, LA 70813, USA}
\affiliation{Dept.~of Physics, University of California, Berkeley, CA 94720, USA}
\affiliation{Lawrence Berkeley National Laboratory, Berkeley, CA 94720, USA}
\affiliation{Institut f\"ur Physik, Humboldt-Universit\"at zu Berlin, D-12489 Berlin, Germany}
\affiliation{Universit\'e Libre de Bruxelles, Science Faculty CP230, B-1050 Brussels, Belgium}
\affiliation{Vrije Universiteit Brussel, Dienst ELEM, B-1050 Brussels, Belgium}
\affiliation{Dept.~of Physics, Chiba University, Chiba 263-8522, Japan}
\affiliation{Dept.~of Physics and Astronomy, University of Canterbury, Private Bag 4800, Christchurch, New Zealand}
\affiliation{Dept.~of Physics, University of Maryland, College Park, MD 20742, USA}
\affiliation{Dept.~of Physics and Center for Cosmology and Astro-Particle Physics, Ohio State University, Columbus, OH 43210, USA}
\affiliation{Dept.~of Astronomy, Ohio State University, Columbus, OH 43210, USA}
\affiliation{Dept.~of Physics, TU Dortmund University, D-44221 Dortmund, Germany}
\affiliation{Dept.~of Subatomic and Radiation Physics, University of Gent, B-9000 Gent, Belgium}
\affiliation{Max-Planck-Institut f\"ur Kernphysik, D-69177 Heidelberg, Germany}
\affiliation{Dept.~of Physics and Astronomy, University of California, Irvine, CA 92697, USA}
\affiliation{Laboratory for High Energy Physics, \'Ecole Polytechnique F\'ed\'erale, CH-1015 Lausanne, Switzerland}
\affiliation{Dept.~of Physics and Astronomy, University of Kansas, Lawrence, KS 66045, USA}
\affiliation{Dept.~of Astronomy, University of Wisconsin, Madison, WI 53706, USA}
\affiliation{Dept.~of Physics, University of Wisconsin, Madison, WI 53706, USA}
\affiliation{Institute of Physics, University of Mainz, Staudinger Weg 7, D-55099 Mainz, Germany}
\affiliation{University of Mons-Hainaut, 7000 Mons, Belgium}
\affiliation{Bartol Research Institute and Department of Physics and Astronomy, University of Delaware, Newark, DE 19716, USA}
\affiliation{Dept.~of Physics, University of Oxford, 1 Keble Road, Oxford OX1 3NP, UK}
\affiliation{Dept.~of Physics, University of Wisconsin, River Falls, WI 54022, USA}
\affiliation{Oskar Klein Centre and Dept.~of Physics, Stockholm University, SE-10691 Stockholm, Sweden}
\affiliation{Dept.~of Astronomy and Astrophysics, Pennsylvania State University, University Park, PA 16802, USA}
\affiliation{Dept.~of Physics, Pennsylvania State University, University Park, PA 16802, USA}
\affiliation{Dept.~of Physics and Astronomy, Uppsala University, Box 516, S-75120 Uppsala, Sweden}
\affiliation{Dept.~of Physics and Astronomy, Utrecht University/SRON, NL-3584 CC Utrecht, The Netherlands}
\affiliation{Dept.~of Physics, University of Wuppertal, D-42119 Wuppertal, Germany}
\affiliation{DESY, D-15735 Zeuthen, Germany}

\author{R.~Abbasi}
\affiliation{Dept.~of Physics, University of Wisconsin, Madison, WI 53706, USA}
\author{Y.~Abdou}
\affiliation{Dept.~of Subatomic and Radiation Physics, University of Gent, B-9000 Gent, Belgium}
\author{M.~Ackermann}
\affiliation{DESY, D-15735 Zeuthen, Germany}
\author{J.~Adams}
\affiliation{Dept.~of Physics and Astronomy, University of Canterbury, Private Bag 4800, Christchurch, New Zealand}
\author{M.~Ahlers}
\affiliation{Dept.~of Physics, University of Oxford, 1 Keble Road, Oxford OX1 3NP, UK}
\author{K.~Andeen}
\affiliation{Dept.~of Physics, University of Wisconsin, Madison, WI 53706, USA}
\author{J.~Auffenberg}
\affiliation{Dept.~of Physics, University of Wuppertal, D-42119 Wuppertal, Germany}
\author{X.~Bai}
\affiliation{Bartol Research Institute and Department of Physics and Astronomy, University of Delaware, Newark, DE 19716, USA}
\author{M.~Baker}
\affiliation{Dept.~of Physics, University of Wisconsin, Madison, WI 53706, USA}
\author{S.~W.~Barwick}
\affiliation{Dept.~of Physics and Astronomy, University of California, Irvine, CA 92697, USA}
% ...

\date{\today}

\collaboration{IceCube Collaboration}
\noaffiliation

\begin{abstract}
A search for muon neutrinos from neutralino annihilations in the Sun has been performed with the
IceCube 22-string neutrino detector using data collected in 104.3 days of live-time in 2007. No excess over the expected
atmospheric background has been observed. Upper limits have been obtained on
the annihilation rate of captured neutralinos in the Sun and converted to limits on the WIMP-proton cross-sections for WIMP masses in the range 250 - 5000 GeV.
These results are the most stringent limits to date on neutralino annihilation in the~Sun.
\end{abstract}

\pacs{95.35.+d, 98.70.Sa, 96.50.S-, 96.50.Vg}

\maketitle
Non-baryonic cold dark matter in the form of weakly interacting
massive particles (WIMPs) is one of the most promising solutions to
the dark matter problem~\cite{dark}. The minimal supersymmetric extension of the Standard Model
(MSSM) provides a natural WIMP candidate in the lightest neutralino $\tilde \chi^{0}_{1}$~\cite{mssm}. This particle is weakly interacting only
and, assuming R-parity conservation, is stable and can therefore
survive today as a relic from the Big Bang. A wide range of neutralino
masses, $m_{\tilde \chi^{0}_{1}}$, from 46~GeV~\cite{pdg} to a few tens of TeV~\cite{high_limit_new} is compatible with
observations and accelerator-based measurements. Within these bounds it is possible to construct
models where the neutralino provides the needed relic dark matter~density.

Relic neutralinos in the galactic halo may become gravitationally trapped in the Sun and accumulate
in its center, where they can annihilate each other, producing standard model particles. These may decay, creating neutrinos which can escape and reach the Earth. The search presented here aims at detecting
neutralino annihilations indirectly by observing an excess of such high energy
neutrinos from the~Sun. Limits on the neutrino flux from the Sun have previously been reported by BAKSAN \cite{baksan}, MACRO \cite{macro}, Super-Kamiokande \cite{superk}, and AMANDA \cite{sunWimp}.

The IceCube detector~\cite{icecube} records Cherenkov light in the ice from relativistic charged particles created in neutrino interactions. In 2007 the detector consisted of an array of 22 vertical strings with 60 Digital Optical Modules (DOMs) each, deployed in the clear Antarctic ice at the South Pole at depths between 1450 m and 2450 m below the ice surface. The vertical spacing between DOMs is 17 m and the horizontal distance between strings is 125 m. Each DOM consists of a pressurized glass sphere containing a 25 cm photomultiplier tube (PMT) and a digitizer board. The PMT waveforms are stored when nearest or next-to-nearest DOMs fire within 1 $\mu$s. The trigger selects time windows when eight DOMs produce waveforms within 5 $\mu$s. The reconstructed first~photon arrival times are used to determine the muon~direction.

The background in the search for neutrinos from the Sun comes from air showers created by cosmic ray interactions in the atmosphere. The showers cause downwards going atmospheric muon events, triggering at several hundred Hz, and atmospheric muon neutrino events, triggering at a few mHz. When the Sun is below the horizon, the neutrino signal can be distinguished from the atmospheric muon background by selecting events with upward-going reconstructed muon tracks.

\begin{figure}[t!]
\includegraphics[width=0.55\textwidth]{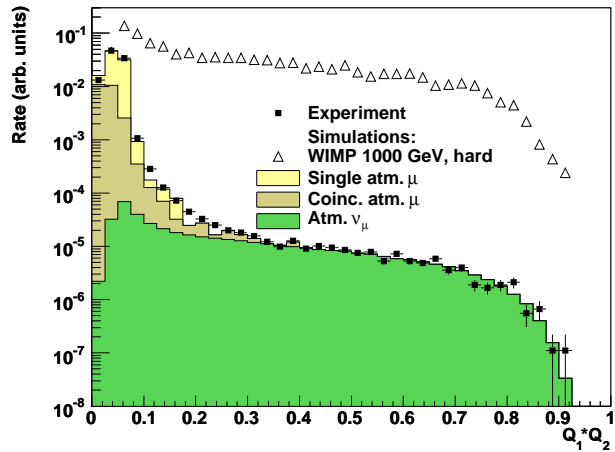}
\caption{\label{fig:q1q2} The product $Q_{1}\times Q_{2}$ of the output values of the two SVMs for the experimental data, a simulated signal ($m_{\tilde \chi^{0}_{1}}$ = 1000 GeV, hard spectrum) and the background. The background has been scaled to match the data rate and it is shown divided into three components: atmospheric neutrinos and single and coincident atmospheric muons.}
\end{figure}

The dataset used in this analysis consists of $\sim4.8$~$\cdot10^{9}$ triggering events taken while the Sun was below the horizon, corresponding to 104.3 days of livetime between June 1st and September 23rd, 2007. The events were processed through several filters to reduce the content of atmospheric muon events and to enrich the dataset in muon-neutrino events. The analysis was performed in a blind manner such that the azimuth of the Sun was not looked at until the selection cuts were~finalized.

Events were first required to have at least ten hit DOMs, and the zenith angle of the line-fit~\cite{reco} first-guess reconstructed track was required to be larger than $70^{\circ}$. Selected events were subjected to Log-Likelihood (LLH) fitting of muon tracks~\cite{reco}, which uses the probability distribution of the photon arrival times. Cuts were then placed on the zenith angle of this reconstruction ($90^{\circ} < \theta_{\mathrm{LLH}}< 120^{\circ}$) and the width of the likelihood optimum ($\sigma_{\mathrm{LLH}} < 10^{\circ}$), to select upwards going events of good quality. Very loose cuts were placed on several kinematic quantities to remove a small number of outlying events.
The final background reduction was then done using Support Vector Machines (SVMs)~\cite{svm}, multi-variate learning machines used to classify events as signal-like or background-like. Twelve event observables, that correlated modestly with one another (correlation coefficient $\lvert c \rvert < 0.5$), were used to train two SVMs with six input observables each. The use of two SVMs allowed minimal correlation ($\lvert c \rvert < 0.3$) between the six observables for each SVM. Training was done with simulated signal events, and a set of real data, not used in the analysis, was taken as background. The observables describe the quality of the track reconstructions and the geometry and the time evolution of the hit pattern, most notably through the opening angle between the line-fit and the LLH tracks, $\sigma_{\mathrm{LLH}}$, the mean minimal distance between the LLH track and the hit DOMs, and the number of hit strings. The SVM input distributions for data and simulated backgrounds were generally in good agreement.

\begin{figure}[t]
\includegraphics[width=0.5\textwidth]{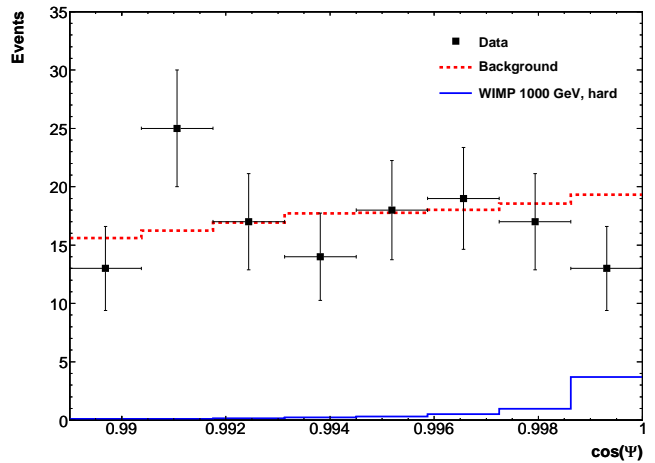}
\caption{\label{fig:cos} Cosine of the angle between the reconstructed track and the direction of the Sun, $\Psi$, for data (squares) with one standard deviation error bars, and the atmospheric background expectation from atmospheric muons and neutrinos (dashed line). Also shown is a simulated signal ($m_{\tilde \chi^{0}_{1}}$ = 1000 GeV, hard spectrum) scaled to $\mu_{s}=6.8$ events (see Table I).}
\end{figure}
Three types of background were simulated: atmospheric muon events from single and coincident air showers were simulated using \texttt{CORSIKA}~\cite{cors}, and atmospheric $\nu_{\mu}$ events were simulated following the Bartol spectrum~\cite{bartol}. Solar-WIMP signals were simulated with \texttt{WimpSim}~\cite{blen}. Two neutralino annihilation channels, $W^{+}W^{-}$ (hard channel) which produces a harder neutrino energy spectrum, and $b\overline{b}$ (soft channel) which gives rise to a softer neutrino energy spectrum, were simulated for five masses $m_{\tilde \chi^{0}_{1}}$ = 250, 500, 1000, 3000, and 5000 GeV. The neutrinos were propagated through the Sun and to the Earth with full flavour oscillation. Absorption in the Sun is important for neutrinos with energies above a few hundred GeV. A muon and a hadronic shower were generated in the ice near the detector. At the vertices the mean energy of simulated signal muons ranges from about 30 GeV to about 150 GeV depending on signal model, see Table~\ref{tab:table1}. For the hard channel $<\!E_{\mu}\!>$ decreases for $m_{\tilde \chi^{0}_{1}}>3$ TeV owing to neutrino absorption in the Sun and secondary neutrino generation. The muon contribution from tau decay was evaluated to be insignificant and tau vertices were therefore neglected. Propagation of muons through the ice was simulated~\cite{mmc}, and the Cherenkov light propagation from the muon to the DOMs was performed with~\cite{pho}, taking into account measured ice properties~\cite{iceprop}.

Fig.~\ref{fig:q1q2} shows the distributions of the product of the two SVM output values, $Q_{1} \times Q_{2}$. As can be seen in the figure the distribution of simulated background is in good agreement with data. The final event sample was selected by requiring $Q_{1} \times Q_{2} > 0.1$. This cut increased the $\mathrm{signal}:\sqrt{\mathrm{background}}$ ratio by a factor of 8.

Simulations predict that the final data sample of 6946 events has an atmospheric $\nu_{\mu}$ event content of 56\%, and that the remainder consists of mis-reconstructed atmospheric muon events. The loose cuts maintain a large effective volume, defined as the detector volume with $100\%$ selection efficiency, since the final signal determination was done on the basis of direction.

After calculating the Sun's position, the observed number of events as a function of the angle to the Sun, $\Psi$, is compared to the atmospheric background expectation in Fig.~\ref{fig:cos}. The angular distribution is consistent with the expected background and no excess of events from the Sun is~observed.

\begin{figure}[t]
\includegraphics[width=0.5\textwidth]{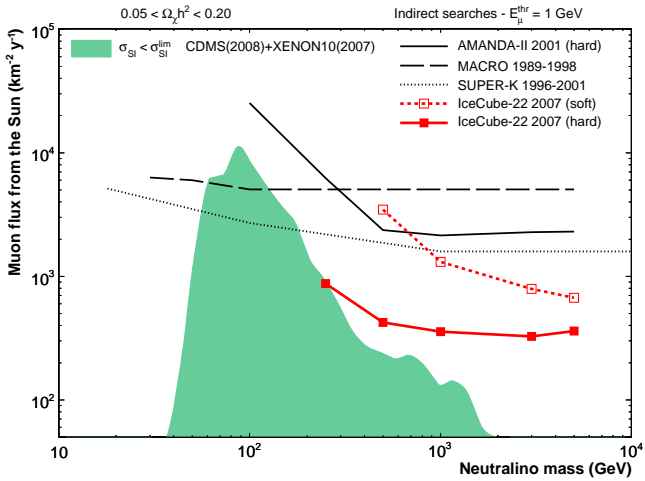}
\caption{\label{fig:flux} Upper limits at the 90\% confidence level on the muon flux from neutralino annihilations in the Sun for the soft ($b\overline{b}$) and hard ($W^{+}W^{-}$) annihilation channels, adjusted for systematic effects, as a function of neutralino mass. The shaded area represents MSSM models not disfavoured by direct searches~\cite{cdms,xenon10}. A muon energy threshold of 1~GeV was used when calculating the flux. Also shown are the limits from MACRO \cite{macro}, Super-K~\cite{superk}, and AMANDA~\cite{sunWimp}.}
\end{figure}

Using likelihood-ratio hypothesis tests the observed $\Psi$ distribution is fitted with a sum of distributions of the simulated signal and the expected background. Here, the expected background is detemined by using real data with randomized azimuth direction of the Sun. We then follow the unified Feldman-Cousins approach~\cite{fc} to construct the confidence intervals on the number of signal events $\mu_{\mathrm{s}}$. The upper 90\% confidence limit ranges between $\mu_{\mathrm{s}}=6.4$ and $\mu_{\mathrm{s}}=8.5$ events depending on signal case, see Table~\ref{tab:table1}.

Simulation studies were used to estimate the systematic uncertainty on the signal effective volume $V_{\mathrm{eff}}$. Uncertainties in the photon propagation in ice and absolute DOM efficiency dominate, contributing $\pm 17\%$ to $\pm 24\%$ depending on the signal model. The total systematic uncertainty on $V_{\mathrm{eff}}$ ranges from $\pm 19\%$ for the highest $m_{\tilde \chi^{0}_{1}}$ to $\pm 26\%$ for the lowest $m_{\tilde \chi^{0}_{1}}$. Deviations in the event rate between data and background simulations are within the systematic uncertainty. These uncertainties are included in the results presented below.

From the upper limits on $\mu_{\mathrm{s}}$ we calculate the limit on the neutrino to muon conversion rate $\Gamma_{\nu\rightarrow\mu} = \frac{\mu_{\mathrm{s}}(\Psi)}{V_{\mathrm{eff}}\cdot t}$, for the livetime $t$. Using the signal simulation~\cite{blen}, we can convert this rate to a limit on the neutralino annihilation rate in the Sun, $\Gamma_{\mathrm{A}}$, see Table~\ref{tab:table1}. Results from different experiments are commonly compared by calculating the limit on the muon flux above 1 GeV, $\Phi_{\mu}$, which is also shown in Table~\ref{tab:table1} together with the sensitivity, $\overline{\Phi}_{\mu}$, the median limit obtained from simulations with no signal. A downward fluctuation in the data close to the position of the Sun results in limits lower than the sensitivity. Within $\Psi < 3^{\circ}$, corresponding to the rightmost bin in Fig.~\ref{fig:cos}, the fluctuation has a probability of 8.8\%. In this bin we expect less than 0.4 background events from solar atmospheric neutrinos~\cite{sanu}.

\begin{figure}[t]
\includegraphics[width=0.5\textwidth]{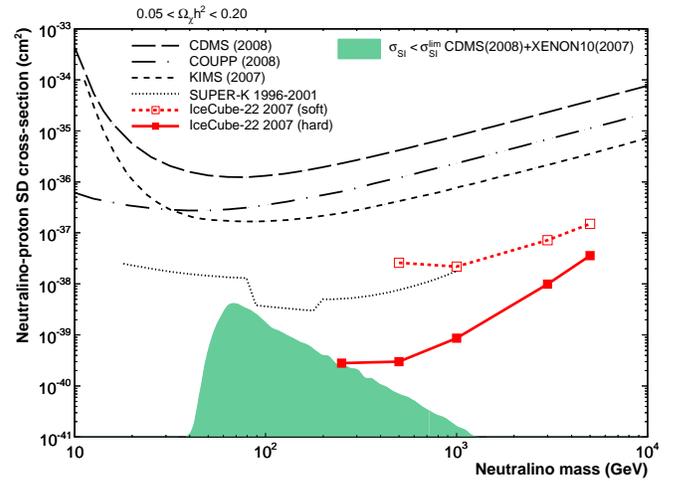}
\caption{\label{fig:xsec} Upper limits at the 90\% confidence level on the spin-dependent neutralino-proton cross-section $\sigma^{\it{SD}}$ for the soft ($b\overline{b}$) and hard ($W^{+}W^{-}$) annihilation channels, adjusted for systematic effects, as a function of neutralino mass. The shaded area represents MSSM models not disfavoured by direct searches~\cite{cdms,xenon10} based on $\sigma^{\it{SI}}$. Also shown are the limits from CDMS~\cite{cdms}, COUPP~\cite{coupp}, KIMS~\cite{kims} and Super-K~\cite{superk}.}
\end{figure}

\begin{table*}
\caption{\label{tab:table1}Upper limits on the number of signal events $\mu_{\mathrm{s}}$, the conversion rate $\Gamma_{\nu\rightarrow\mu}$, the neutralino annihilation rate in the Sun $\Gamma_{\mathrm{A}}$, the muon flux $\Phi_{\mu}$, and the neutralino-proton scattering cross-sections (spin-independent, $\sigma^{\it{SI}}$, and spin-dependent, $\sigma^{\it{SD}}$), at the 90\% confidence level including systematic errors. The sensitivity $\overline{\Phi}_{\mu}$ (see text) is shown for comparison. Also shown is the median angular error $\Theta$, the mean muon energy $<\!E_{\mu}\!>$, the effective volume $V_{\mathrm{eff}}$, and the $\nu_{\mu}$ effective area $A_{\mathrm{eff}}$.}
\begin{ruledtabular}
 \begin{tabular}{cc|cccc|c|cc|cccc}
 $m_{\tilde \chi^{0}_{1}}$&Channel&$\mu_{\mathrm{s}}$& $\Gamma_{\nu\rightarrow\mu}$&$\Gamma_{\mathrm{A}}$&$\Phi_{\mu}$&$\overline{\Phi}_{\mu}$&$\sigma^{\it{SI}}$&$\sigma^{\it{SD}}$&$\Theta$&$<\!E_{\mu}\!>$&$V_{\mathrm{eff}}$&$A_{\mathrm{eff}}$\\
 (GeV)&  &  & ($\mathrm{km}^{-3} \mathrm{y}^{-1}$)& ($s^{-1}$) &($\mathrm{km}^{-2} \mathrm{y}^{-1}$)& ($\mathrm{km}^{-2} \mathrm{y}^{-1}$)&($\mathrm{cm}^{2}$)&($\mathrm{cm}^{2}$)& &(GeV) & ($\mathrm{km}^{3}$)& ($\mathrm{m}^{2}$)\\\hline
250 &Hard& 7.5  & $3.2\cdot 10^{3}$& $6.0\cdot 10^{21}$& $8.8 \cdot 10^{2}$& $1.6\cdot 10^{3}$& $3.7\cdot 10^{-43}$ & $2.8\cdot 10^{-40}$&$3.2^{\circ}$ & 68.7  & $8.1\cdot 10^{-3}$&$1.3\cdot10^{-4}$\\ \hline
500 &Soft& 8.5  & $2.8\cdot 10^{4}$& $1.4\cdot 10^{23}$& $3.5 \cdot 10^{3}$& $5.7\cdot 10^{3}$& $2.5\cdot 10^{-41}$ & $2.6\cdot 10^{-38}$&$3.5^{\circ}$ & 28.8  & $1.1\cdot 10^{-3}$&$6.7\cdot10^{-6}$\\
    &Hard& 6.8  & $1.0\cdot 10^{3}$& $1.6\cdot 10^{21}$& $4.2 \cdot 10^{2}$& $7.9\cdot 10^{2}$& $2.9\cdot 10^{-43}$ & $3.0\cdot 10^{-40}$&$2.9^{\circ}$ & 111   & $2.4\cdot 10^{-2}$&$4.9\cdot10^{-4}$\\ \hline
1000&Soft& 7.5  & $7.8\cdot 10^{3}$& $3.0\cdot 10^{22}$& $1.3 \cdot 10^{3}$& $2.4\cdot 10^{3}$& $1.8\cdot 10^{-41}$ & $2.2\cdot 10^{-38}$&$3.2^{\circ}$ & 40.8  & $3.4\cdot 10^{-3}$&$2.6\cdot10^{-5}$\\
    &Hard& 6.8  & $6.7\cdot 10^{2}$& $1.2\cdot 10^{21}$& $3.6 \cdot 10^{2}$& $6.3\cdot 10^{2}$& $7.2\cdot 10^{-43}$ & $8.7\cdot 10^{-40}$&$2.9^{\circ}$ & 146   & $3.5\cdot 10^{-2}$&$7.6\cdot10^{-4}$\\ \hline
3000&Soft& 7.8  & $3.5\cdot 10^{3}$& $1.1\cdot 10^{22}$& $7.9 \cdot 10^{2}$& $1.3\cdot 10^{3}$& $5.3\cdot 10^{-41}$ & $7.2\cdot 10^{-38}$&$3.1^{\circ}$ & 55.8  & $7.7\cdot 10^{-3}$&$6.9\cdot10^{-5}$\\
    &Hard& 6.4  & $6.1\cdot 10^{2}$& $1.5\cdot 10^{21}$& $3.3 \cdot 10^{2}$& $6.1\cdot 10^{2}$& $7.4\cdot 10^{-42}$ & $9.9\cdot 10^{-39}$&$2.9^{\circ}$ & 149   & $3.7\cdot 10^{-2}$&$7.3\cdot10^{-4}$\\ \hline
5000&Soft& 7.5  & $2.8\cdot 10^{3}$& $8.3\cdot 10^{21}$& $6.7 \cdot 10^{2}$& $1.1\cdot 10^{3}$& $1.1\cdot 10^{-40}$ & $1.5\cdot 10^{-37}$&$3.1^{\circ}$ & 59.9  & $9.3\cdot 10^{-3}$&$8.6\cdot10^{-5}$\\
    &Hard& 6.8  & $7.0\cdot 10^{2}$& $2.0\cdot 10^{21}$& $3.6 \cdot 10^{2}$& $6.6\cdot 10^{2}$& $2.6\cdot 10^{-41}$ & $3.6\cdot 10^{-38}$&$2.9^{\circ}$ & 142   & $3.4\cdot 10^{-2}$&$6.2\cdot10^{-4}$\\
\end{tabular}
\end{ruledtabular}
\end{table*} 

The 90\% confidence upper limit on $\Phi_{\mu}$ as a function of $m_{\tilde \chi^{0}_{1}}$ is shown in Fig.~\ref{fig:flux}, compared to other limits \cite{macro,superk,sunWimp}, and MSSM model predictions~\cite{dsusy}. In the plot, the shaded area represents neutralino models not disfavoured by the direct detection experiments CDMS~\cite{cdms} and XENON-10~\cite{xenon10}, based on their limit on the spin-independent neutralino-proton cross-section.

The limits on the annihilation rate can be converted into limits on the spin-dependent, $\sigma^{\it{SD}}$, and 
spin-independent, $\sigma^{\it{SI}}$, neutralino-proton cross-sections, allowing a more direct comparison with the results of direct search experiments. Since capture in the Sun is dominated by $\sigma^{\it{SD}}$, indirect searches are expected to be competitive in setting limits on this quantity. Assuming equilibrium between the capture and annihilation rates in the Sun, the annihilation rate is directly proportional to the cross-section. A limit on $\sigma^{\it{SD}}$ is found by setting $\sigma^{\it{SI}}$ to zero, and vice versa. We have used DarkSUSY~\cite{dsusy} and the method described in \cite{conv} to perform the conversion. The results are shown in Table~\ref{tab:table1}. We assumed a local WIMP density of $0.3~\mathrm{GeV/cm}^{3}$, and a Maxwellian WIMP velocity distribution with a dispersion of 270~km/s. Planetary effects on the capture were neglected. Fig.~\ref{fig:xsec} shows the IceCube-22 limits on $\sigma^{\it{SD}}$ compared with other bounds \cite{cdms,coupp,kims,superk}, and the MSSM model space defined as for Fig.~\ref{fig:flux}. Indirect searches for dark matter in the Sun complement direct searches on Earth in several respects. WIMPs in the Sun would accumulate over a long period and therefore sample over different dark matter densities in the galactic halo. This gravitational accumulation is sensitive to low WIMP velocities while direct detection recoil experiments are more sensitive at higher velocities.

In conclusion, we have presented the most stringent limits to date on neutralino annihilations in the Sun, improving on the 2001 AMANDA~\cite{sunWimp} limits by at least a factor of six for hard channels. We also present the most stringent limits on the spin-dependent WIMP-proton cross-section for neutralino masses above 250 GeV. The full IceCube detector with the DeepCore extension~\cite{dc} is expected to test viable MSSM models down to 50 GeV.

We acknowledge support from the following agencies:
U.S. National Science Foundation-Office of Polar Programs,
U.S. National Science Foundation-Physics Division,
U. of Wisconsin Alumni Research Foundation,
U.S. Department of Energy, NERSC,
the LONI grid;
Swedish Research Council,
K.~\&~A.~Wallenberg Foundation, Sweden;
German Ministry for Education and Research,
Deutsche Forschungsgemeinschaft;
Fund for Scientific Research,
IWT-Flanders,
BELSPO, Belgium;
the Netherlands Organisation for Scientific Research;
M.~Ribordy is supported by SNF (Switzerland);
A.~Kappes and A.~Gro{\ss} are supported by the EU Marie Curie OIF Program.
We thank J.~Edsj\"{o} for DarkSUSY support.

\end{document}